\documentclass[conference]{IEEEtran}
\IEEEoverridecommandlockouts
\usepackage{cite}
\usepackage{amsmath,amssymb,amsfonts}
\usepackage{algorithmic}
\usepackage{graphicx}
\usepackage{textcomp}
\usepackage{xcolor}

\def\BibTeX{{\rm B\kern-.05em{\sc i\kern-.025em b}\kern-.08em
    T\kern-.1667em\lower.7ex\hbox{E}\kern-.125emX}}

\usepackage{listings}
\usepackage{lstlangcoq}
\usepackage{color}

\definecolor{dkgreen}{rgb}{0,0.6,0}
\definecolor{gray}{rgb}{0.5,0.5,0.5}
\definecolor{mauve}{rgb}{0.58,0,0.82}

\usepackage{hyperref}

\begin{document}

\title{PubSub implementation in Haskell with formal verification in Coq}

\author{
    \IEEEauthorblockN{Boro Sitnikovski\IEEEauthorrefmark{1}, Biljana Stojcevska\IEEEauthorrefmark{1}, Lidija Goracinova-Ilieva\IEEEauthorrefmark{1}, Irena Stojmenovska\IEEEauthorrefmark{2}}
    \IEEEauthorblockA{\IEEEauthorrefmark{1}Faculty of Informatics, UTMS Skopje
    \\\ buritomath@gmail.com, \{b.stojcevska, l.goracinova\}@utms.edu.mk}
    \IEEEauthorblockA{\IEEEauthorrefmark{2}School of Computer Science and IT, UACS Skopje
    \\\ irena.stojmenovska@uacs.edu.mk}
}

\maketitle

\thispagestyle{plain}
\pagestyle{plain}

\begin{abstract}
In the cloud, the technology is used on-demand without the need to install anything on the desktop. Software as a Service is one of the many cloud architectures. The PubSub messaging pattern is a cloud-based Software as a Service solution used in complex systems, especially in the notifications part where there is a need to send a message from one unit to another single unit or multiple units. Haskell is a generic typed programming language which has pioneered several advanced programming language features. Based on the lambda calculus system, it belongs to the family of functional programming languages. Coq, also based on a stricter version of lambda calculus, is a programming language that has a more advanced type system than Haskell and is mainly used for theorem proving i.e. proving software correctness. This paper aims to show how PubSub can be used in conjunction with cloud computing (Software as a Service), as well as to present an example implementation in Haskell and proof of correctness in Coq.
\end{abstract}


\section{Introduction}

A cloud can be both software and infrastructure. It can be an application accessed via the Internet or a server provided when needed. If a service can be accessed by a device, regardless of the operating system of that device, then that service is cloud-based \cite{b1}.

Typically, three criteria are defined as labels whether a particular service is a cloud service \cite{b1}:

\begin{itemize}
\item the service is available through a web browser or web services API,
\item zero capital spending is needed to get started,
\item payment is required only for those services that are used.
\end{itemize}

The PubSub (publish-subscribe) pattern allows for easy message transfer to specific channels.

Haskell and Coq are programming languages designed with an aim to accomplish software correctness, based on a typed version of the lambda calculus system \cite{b2}, \cite{b3}, \cite{b4}.

\section{Architecture}

\subsection{Cloud computing}

We can categorize most cloud-computing projects in three basic categories:

\begin{itemize}
\item projects that deploy services for multiple applications or clients,
\item projects that are single, standalone cloud applications,
\item cloud-based service provider projects (e.g. Google).
\end{itemize}

Besides these project categories, there are three main cloud-based architectures \cite{b5}:

\begin{itemize}
\item SaaS - software as a service (e.g. Google Apps, Salesforce, Dropbox, games such as World of Warcraft)
\item PaaS - platform as a service (e.g. Windows Azure, Heroku, Google App Engine, WordPress web site development platform)
\item IaaS - Infrastructure as a Service (e.g. Google Cloud Platform, Amazon Web Services, Microsoft Azure, etc.)
\end{itemize}

\subsection{Software as a Service}

\begin{figure}[htbp]
\centerline{\includegraphics[width=0.40\textwidth]{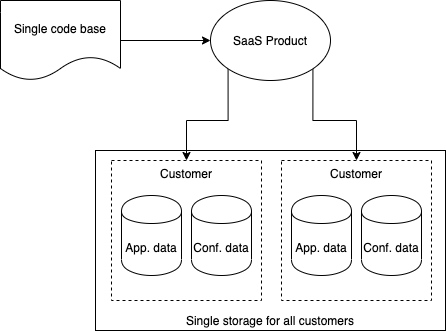}}
\caption{SaaS architecture}
\label{fig1}
\end{figure}

SaaS refers to software hosted on the cloud in a central location (Figure ~\ref{fig1}) \cite{b1}. Typically, this architecture consists of web-based software, but it is not limited to.

SaaS applications are accessed through a client such as a web browser \cite{b1}. This architecture applies to many business applications, including Enterprise Resource Planning (ERP), Customer Relationship Management (CRM), Office software, messaging software, etc \cite{b5}. SaaS is involved in the strategy of almost all leading software companies - Amazon, Google, Microsoft \cite{b5}.

\subsection{PubSub service}

The PubSub pattern is a messaging pattern where publishers can send messages to specific channels to which subscribers are subscribed \cite{b6}. The commands \texttt{SUBSCRIBE}, \texttt{UNSUBSCRIBE}, and \texttt{PUBLISH} implement this pattern. This separation of publishers and subscribers provides better scalability to the service \cite{b6}.

\begin{figure}[htbp]
\centerline{\includegraphics[width=0.40\textwidth]{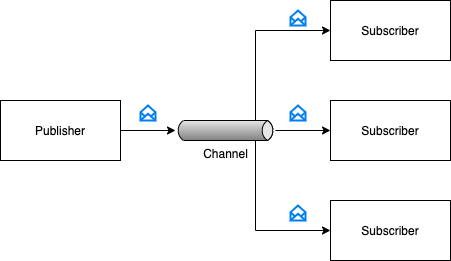}}
\caption{PubSub architecture}
\label{fig2}
\end{figure}

There are three basic components to understanding a PubSub messaging scheme, as shown in Fig. ~\ref{fig2}:

\begin{itemize}
\item Publisher - Publishes messages to the communication infrastructure
\item Subscriber - Subscribes to a specific channel/category of messages
\item Information Infrastructure (Channel) - Handles subscriptions and receives publisher's messages
\end{itemize}

\section{Implementation in Haskell}

In our implementation of PubSub, we use the Haskell programming language. Haskell is an advanced functional programming language. The development of Haskell is rooted in mathematics and computer science research \cite{b2}.

\subsection{Subscription module}

This module represents the business logic functions of the PubSub architecture. Namely, each subscription is represented as a pair of \texttt{Channel} and \texttt{Connection}. Further, all such subscriptions are merged into a single list of subscriptions.

The function \texttt{getHandlesByCh} takes a \texttt{Channel} and a list of \texttt{Subscription}s and then returns a filtered list of subscriptions such that the channel is matched. The Haskell's built-in function \texttt{filter} \cite{b2} will be used by the \texttt{publish} command.

\begin{lstlisting}
getHandlesByCh :: Channel -> [Subscription a] -> [Subscription a]
getHandlesByCh c = filter (\x -> c == fst x)
\end{lstlisting}

The function \texttt{removeHandleByCon} takes a connection and a list of \texttt{Subscription}s and then returns a filtered list of subscriptions such that the selected connection is not contained. This will be used by the \texttt{quit} command.

\begin{lstlisting}
removeHandleByCon :: (Eq a) => Connection a -> [Subscription a] -> [Subscription a]
removeHandleByCon h = filter (\x -> h /= snd x)
\end{lstlisting}

The function \texttt{removeSubscription} takes a channel and a connection and a list of \texttt{Subscription}s and then returns a filtered list of subscriptions such that the selected connection and channel are not contained. This will be used by the \texttt{unsubscribe} command.

\begin{lstlisting}
removeSubscription :: (Eq a) => Channel -> Connection a -> [Subscription a] -> [Subscription a]
removeSubscription c h = filter (\(x, y) -> not (x == c && y == h))
\end{lstlisting}

The function \texttt{addSubscription} takes a channel and a connection and a list of \texttt{Subscription}s and then returns a list such that the selected connection and channel are contained. This will be used by the \texttt{subscribe} command.

\begin{lstlisting}
addSubscription :: Channel -> Connection a -> [Subscription a] -> [Subscription a]
addSubscription c h s = (c, h) : s
\end{lstlisting}

The function \texttt{hInSubscription} checks if a given channel/subscription is contained into a list of subscriptions.

\begin{lstlisting}
hInSubscription :: (Eq a) => Channel -> Connection a -> [Subscription a] -> Bool
hInSubscription c h = any (\(x, y) -> x == c && y == h)
\end{lstlisting}

The next function is a helper function for the \texttt{subscribe} command.

\begin{lstlisting}
subscribe ch h s = if not (hInSubscription ch h s) then addSubscription ch h s else s
\end{lstlisting}

Conversely, the following function is for the \texttt{unsubscribe} command.

\begin{lstlisting}
unsubscribe ch h s = if hInSubscription ch h s then removeSubscription ch h s else s
\end{lstlisting}

Finally, the last function \texttt{publish} uses the condition the types \texttt{t} and \texttt{m} to be \texttt{Foldable} (to be iterated) and \texttt{Monad} (to be chain-operated, for example in IO) respectively. This covers a more general case.

However, in this specific case (PubSub), \texttt{mapM} iterates through the list of subscriptions and perform the function \texttt{f} on each subscription individually. In the PubSub implementation, \texttt{f} is a function that writes to an IO object (\texttt{hPutStrLn}). This generalization, even-though complex, is necessary to easily prove correctness, since in Coq, IO does not exist as a concept.

\begin{lstlisting}
publish :: (Traversable t, Monad m) => (t2 -> m b) -> t (a, t2) -> m (t b)
publish f = Data.Traversable.mapM (\(_, y) -> f y)
\end{lstlisting}

\subsection{Main module}

This module represents the entry-point of the program, where the logic to accept new clients is initialized.

\begin{figure}[htbp]
\centerline{\includegraphics[width=0.3\textwidth]{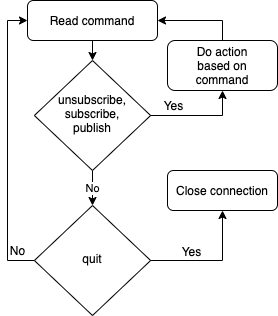}}
\caption{PubSub algorithm}
\label{fig3}
\end{figure}

The main module accepts new connections indefinitely, launching a new thread for every connection. These threads use the algorithm described in Fig. ~\ref{fig3}.

The algorithm handles different commands passed through the communication channel. For a given command, list of subscriptions, and a connection, the algorithm updates the list of subscriptions and returns a boolean result, representing the success of the command's execution. Accepted commands are \texttt{unsubscribe}, \texttt{subscribe}, \texttt{publish} and \texttt{quit}. This algorithm will be executed recursively until the client closes the connection.

\subsection{Running example}

In this subsection, we will demonstrate the usage of our application through the client program \texttt{telnet}. We will create two connections, subscribed to channel 1 and 2 respectively, and then we will send messages to these subscribers through a third connection.

First connection:

\begin{lstlisting}[language=sh]
$ telnet localhost 123 import
Connected to localhost.
Write 'publish <ch> <msg>' to publish, 'subscribe <ch>' to subscribe.
> subscribe 1
\end{lstlisting}

Second connection:

\begin{lstlisting}[language=sh]
$ telnet localhost 123
Connected to localhost.
Write 'publish <ch> <msg>' to publish, 'subscribe <ch>' to subscribe.
> subscribe 2
\end{lstlisting}

Third connection:

\begin{lstlisting}[language=sh]
$ telnet localhost 123
Connected to localhost.
Escape character is '^]'.
Write 'publish <ch> <msg>' to publish, 'subscribe <ch>' to subscribe.
> publish 1 hey there
> publish 2 hello!
\end{lstlisting}

The first and the second connection will receive \texttt{[1] hey there} and \texttt{[2] hello!} respectively.

\section{Formal proof of correctness in Coq}

Coq is a programming language used as an interactive theorem prover that enables expressing mathematical definitions \cite{b3}. By mechanically validating evidence of mathematical claims, Coq helps in producing a certified program. It is based on the theory of Calculus of Inductive Constructions, a system in the lambda calculus family \cite{b4}. Haskell is also based on a weaker version of lambda calculus, hence there is a strong connection between these two programming languages.

Coq's initial release was done in 1989 by INRIA \cite{b3}. This programming language has support for representing dependent types \cite{b7}. These types are precisely what enables us to write mathematical proofs \cite{b8}.

Haskell has no support for dependent types, so the first challenge that we face is how to prove some basic property of our programming code (written in Haskell) in Coq. For this, we use an existing tool called hs-to-coq \cite{b9} that allows us to convert Haskell code to Coq. The reverse conversion is already supported in Coq itself. Using this tool, we get the \texttt{Subscription.v} (Coq source code) file.

Semantically, this auto-generated file from hs-to-coq contains the same functions that we have defined earlier in Haskell. The only difference is the syntax, where the Coq syntax is used instead of Haskell. At this point, we can start using the power of Coq.

Thus, we create the following \texttt{Proofs.v} file that has the following content:

\begin{lstlisting}[language=Coq]
Require Import Prelude.
Require Import Subscription.
Require Import Proofs.GHC.Base.
Require Import Data.Semigroup.
\end{lstlisting}

Now we can finally work on the proof itself. We will prove a few simple properties for this paper's argument. We first define a list of subscriptions to use in our proofs:

\begin{lstlisting}[language=Coq]
Definition subs := addSubscription #1 &"fh01" nil.
\end{lstlisting}

We then proceed showing that

\(\texttt{length}( \texttt{getHandlesByCh} \ 1 \ \texttt{subs} ) = 1\)

\begin{lstlisting}[language=Coq]
Lemma example_1 : GHC.List.length (getHandlesByCh #1 subs) = 1%Z.
Proof.
  auto.
Qed.
\end{lstlisting}

We will explain the syntax used in the proof briefly. The interested reader can look at the details in \cite{b10}.

In the code above we have proved the lemma named \texttt{example\_1} which states that the length of the evaluation of \texttt{getHandlesByCh 1 subs} is 1. We begin the proof using the \texttt{Proof.} command. What follows afterward are commands called tactics. These are macro commands that use the lambda calculus rules to simplify formulas. In this case, by using the \texttt{auto} tactic, Coq can mechanically prove the claim, because the claim is simple enough. If the claim was a bit more complex, we would have to use different tactics.

We run Coq and it executes the code mechanically, returning no errors meaning the proof is complete.

\begin{lstlisting}[language=Coq]
Lemma example_2 : GHC.List.length (getHandlesByCh #2 subs) = 0%Z.
Proof.
  auto.
Qed.
\end{lstlisting}

The lemma \texttt{example\_2} is similar to \texttt{example\_1}. It claims that the subscription of channel 2 does not exist in the list, namely that the length of such a list is 0.

\begin{lstlisting}[language=Coq]
Lemma example_3 : getOption (publish (fun y => GHC.Base.return_ 1) subs) = Some (1 :: nil).
Proof.
  auto.
Qed.
\end{lstlisting}

The lemmas \texttt{example\_3} and \texttt{example\_4} prove that the second argument of \texttt{publish} has an effect on the output of the subscriptions. This is as expected, since this is how we defined \texttt{publish} in Haskell earlier.

\begin{lstlisting}[language=Coq]
Lemma example_4 : getOption (publish (fun y => GHC.Base.return_ y) subs) = Some (&"fh01" :: nil).
Proof.
  auto.
Qed.
\end{lstlisting}

With the following lemma we will prove the fact:

\(\forall l, \texttt{hInSub}(1, \texttt{"fh01"}, (\texttt{addSub}, 1, \texttt{"fh01"}, l)\)

That is, for all lists \texttt{l}, upon which a subscription 1 with \texttt{"fh01"} is added, the function \texttt{hInSubscription} returns true.

We further take a brief look at the kinds of errors that Coq may return.

\begin{lstlisting}[language=Coq]
Lemma example_5 : forall l : list (Subscription Base.String), hInSubscription #1 &"fh01" (addSubscription #1 &"fh01" l) = true.
Proof.
Qed.
\end{lstlisting}

The example given above results with the following error from Coq:

\vfill\break 

\begin{lstlisting}[language=sh]
Error:  (in proof example_5): Attempt to save an incomplete proof
\end{lstlisting}

It warns us that the proof is not complete. We can use the command \texttt{Show Existentials} to see the current state of the proof:

\begin{lstlisting}[language=sh]
Existential 1 =
?Goal : [
        |- forall l : list (Subscription String),
           hInSubscription 1%Z &"fh01"
             (addSubscription 1%Z
                &"fh01" l) =
           true]
\end{lstlisting}

This proof is also simple enough for Coq, so we can use \texttt{auto} to complete it as well.

\section*{Conclusions}

Moving to the cloud is one of the current challenges in enterprises. This technology provides a new "on-demand" paradigm for information and communication technologies.

The advantages are cheaper systems, access from countless devices, centralization of data. The main disadvantage is that an Internet connection is required.

PubSub is just one of the many SaaS possibilities. As the usage of such systems continues to rise, formally proving correctness according to specifications is crucial to the system functioning and accuracy as expected. The solution presented in this paper demonstrates how Haskell in conjunction with Coq can be applied to perform this vital step in the process of cloud-based software development.

\end{document}